\documentclass[journal,10pt]{IEEEtran}

\ifCLASSINFOpdf
\else
   \usepackage[dvips]{graphicx}
\fi
\usepackage{url}
\usepackage{subcaption}
\usepackage{cite}
\usepackage{graphicx}
\usepackage{amsmath}
\usepackage{amsthm}
\usepackage{comment}
\usepackage{xcolor}
\usepackage{amsfonts}
\usepackage{bm}
\usepackage{booktabs}
\usepackage[hidelinks]{hyperref}
\usepackage{footmisc}

\begin{document}

\title{Noise Morphing for Audio Time Stretching} 

\author{Eloi Moliner$^\ast$, Leonardo Fierro$^\ast$, Alec Wright, Matti S. Hämäläinen, and Vesa Välimäki, \IEEEmembership{Fellow, IEEE}
\thanks{$\ast$ E. Moliner and L. Fierro contributed equally to this work.}
\thanks{To be submitted for peer-review on 21.12.2023. This work has been financed in part by Nokia Technologies through the DeepSlow 2 project (Aalto University project no. 400610). L. Fierro’s work has been supported by the Aalto ELEC Doctoral School. This work is part of the activities of the NordicSMC network (NordForsk project no. 86892). }
\thanks{E. Moliner, L. Fierro, A. Wright and V. Välimäki are with the Acoustics Lab, Department of Information and Communications Engineering, Aalto University, Espoo, Finland (e-mail: name.surname@aalto.fi).}
\thanks{M. S. Hämäläinen is with Nokia Technologies, Tampere, Finland (e-mail: matti.s.hamalainen@nokia.fi).}}

\markboth{ VOL. X, NO. X, DECEMBER 2023}
{Shell \MakeLowercase{\textit{et al.}}: Bare Demo of IEEEtran.cls for IEEE Journals}
\maketitle

\begin{abstract}
This letter introduces an innovative method to enhance the quality of audio time stretching by precisely decomposing a sound into sines, transients, and noise and by improving the processing of the latter component.
While there are established methods for time-stretching sines and transients with high quality, the manipulation of noise or residual components has lacked robust solutions in prior research. 
The proposed method combines sound decomposition with previous techniques for audio spectral resynthesis. 
The time-stretched noise component is achieved by morphing its time-interpolated spectral magnitude with a white-noise excitation signal. This method stands out for its simplicity, efficiency, and audio quality. The results of a subjective experiment affirm the superiority of this approach over current state-of-the-art methods across all evaluated stretch factors. The proposed technique notably excels in extreme stretching scenarios, signifying a substantial elevation in performance.
The proposed method holds promise for a wide range of applications in slow-motion media content, such as music or sports video production.


\end{abstract}

\begin{IEEEkeywords}
Audio systems, interpolation, signal restoration, spectral analysis, timbre.
\end{IEEEkeywords}

\IEEEpeerreviewmaketitle


\section{Introduction}

Audio time-scale modification (TSM), a critical process in audio signal processing, involves adjusting the temporal duration of a sound signal without altering its pitch \cite{moulines1990pitch, bonada2000automatic, driedger2016review, damskagg2017audio}. This operation is integral in various applications, such as music production \cite{cliff2000hang}, sound design \cite{Valimaki2018, malloy2022timbral}, and multimedia content manipulation \cite{moinet2013slowdio, roberts2021deep}. This task becomes especially challenging with large stretching factors, where conventional methods, such as the phase vocoder, often introduce perceptual artifacts, e.g., transient smearing, loss of presence, and phasiness \cite{laroche1997phase, driedger2016review, damskagg2017audio}.
The subjective nature of audio time stretching further complicates the problem, as there is no clear objective metric for evaluation \cite{fierro2020towards, roberts2021deep}. The inherently ill-defined nature of this task, as there is no ideal reference signal, is shaped by subjective expectations and perceptual nuances. 

The best performing TSM methods apply the Short-Time Fourier Transform (STFT), manipulate the spectrogram of the signal to change its duration, and then apply the inverse STFT to reconstruct the time-scaled signal \cite{roebel2010shape, driedger2016review, damskagg2017audio}. Established TSM methods have predominantly focused on the separation and accurate manipulation of sinusoidal and transient components of sounds \cite{driedger2013improving,   driedger2014tsm, roma2019time}. The noise component describes sound nuances and textures, e.g. plucking or bowing noise from stringed instruments, and is often the main descriptor for environmental sounds \cite{liao2012stretching, fierro2023extreme}. Common TSM approaches, including phase vocoder-based methods, struggle to provide precise descriptions and scaling for such sound nuances, compromising the final time-stretched audio quality \cite{liao2012stretching, timefrequency2011dafx}. The use of a three-way decomposition to isolate the noise component from sines and transients \cite{verma1998analysis, levine1998sines+, verma1998time, fierro2023enhanced}, in combination with phase randomization \cite{serra1990spectral, hanna2003time} in the resynthesis process, showed a first improvement in the quality of the stretched noise component \cite{damskagg2017audio, fierro2023enhanced}. A solution involving a Wavenet neural synthesizer for the noise component has also proved successful for extreme time stretching of environmental sounds \cite{fierro2023extreme}. 

Previous solutions targeting time-stretching of real-world sounds modeled the stretched noise component via linear interpolation of white Gaussian noise, with the spectral magnitude of the original sound around detected transients \cite{moinet2013audio, moinet2013slowdio}, or with the residual component of the original sound after the sines were removed \cite{apel2014sinusoidality}. These solution compromise the audio quality when applied to general sounds as  they are designed for noisy signals and do not feature a three-way decomposition for transient handling. 
An alternative technique leveraged generative adversarial networks for TSM of speech signals \cite{cohen2022speech}, but its data-driven nature imposes limitations on its application to general audio.

This letter introduces ``Noise Morphing" (NM), an approach that combines the core idea behind the aforementioned techniques and the sines-transients-noise decomposition (STN). This involves producing a white-noise excitation signal of equal length to the output signal of the TSM processing. The white-noise signal is morphed with interpolated log-magnitude spectra of the noise component extracted from the target signal. The novelty lies in the application of spectral morphing within the STN framework, which adds a new layer of precision to the TSM processing chain: in the proposed approach, each of the three components is individually processed with the most suitable technique, before being recombined into a time-stretched mixture \cite{verma2000extending,fierro2023enhanced}.

This letter is structured as follows. Section~\ref{s:bg} describes the STN decomposition and TSM principles that this work builds upon. Section~\ref{s:nm} details the proposed NM technique. Section~\ref{s:eval} reports the methods and results of a subjective evaluation conducted against several other TSM algorithms to validate the effectiveness of the novel approach. Section~\ref{s:end} concludes the letter.

\section{Background}\label{s:bg} 


According to the STN model \cite{verma1998analysis, fierro2023enhanced}, any sound can be described as the summation of tonal content (sines), impulsive events (transients), and sound nuances (noise). In this letter, audio signals are decomposed into these three components via soft spectral masking of their spectrograms, which leads to a fuzzy decomposition with perfect reconstruction and is the best method to date for this specific task \cite{fierro2023enhanced}.

Given an audio signal $\mathbf{x} \in \mathbb{R}^N$, and its Short-Time Fourier Transform (STFT) $\mathbf{X} \in \mathbb{C}^{M\times K}$, one can obtain a set of class masks following  the methodology of Fitzgerald 
\cite{fitzgerald2010harmonic}.
A median filter is applied to the magnitude spectrogram $|\mathbf{X}|$  in the time and frequency directions, and is used to retrieve the tonalness $\mathbf{R}_\textrm{s} \in \mathbb{R}^{M\times K}$ and transientness $\mathbf{R}_\textrm{t} \in \mathbb{R}^{M\times K}$, respectively. Soft masks are then computed as follows \cite{fierro2023enhanced}:
\begin{align}
\mathbf{S}= f&\left(\mathbf{R}_\textrm{s}\right), 
\label{Smask} \\
\mathbf{T} = f&\left(\mathbf{R}_\textrm{t}\right), \label{Tmask} \\
\mathbf{N} = 1&-\mathbf{S}-\mathbf{T}, 
\label{Nmask}
\end{align}
\noindent where $f(a)$ is an element-wise saturating function \cite{fierro2023enhanced}:
\begin{equation}
\begin{aligned}
f(&a) = 
\begin{cases} 
1, & \mbox{if } a \geq \beta_\textrm{U} \\
\sin^2{\Big( \dfrac{\pi}{2} \dfrac{a -\beta_\textrm{L}}{\beta_\textrm{U}-\beta_\textrm{L}}} \Big), & \mbox{if } \beta_\textrm{L} \leq a < \beta_\textrm{U} \\
0, & \mbox{otherwise}.
\end{cases}
\end{aligned}
\end{equation}

The masks \eqref{Smask},  \eqref{Tmask}, and  \eqref{Nmask} are then imposed onto complex spectrogram $\mathbf{X}$ 
via element-wise multiplication to perform the decomposition into the three components. The process is repeated for two consecutive stages using different analysis window lengths and separation factors $\beta_\textrm{U}$ and $\beta_\textrm{L}$ to improve the decomposition quality \cite{tachibana2013singing, driedger2014extending, fierro2023enhanced}. The first stage extracts the sines from the transient and noise residual mixture, using a large analysis window and $\beta_\textrm{U}$ = 0.80 and $\beta_\textrm{L}$ = 0.70 for better frequency resolution; the second uses a short analysis window for better temporal resolution, separating the residual into transients and noise \cite{fierro2023enhanced}, using $\beta_\textrm{U}$ = 0.85 and $\beta_\textrm{L}$ = 0.75. Thus, three spectrogram representations are obtained, one for each component. As a consequence of the fuzzy classification, each time-frequency bin can belong to two classes simultaneously: to the sine and noise classes or to the transient and noise classes \cite{fierro2023enhanced}.

After performing the STN decomposition, different TSM algorithms can be applied for each individual component. The sines are time-stretched using a phase vocoder with identity phase locking \cite{laroche1999improved}, as this has been found successful in previous studies \cite{damskagg2017audio, roberts2021deep, fierro2023enhanced, fierro2023extreme}. 
Transients are preserved after extraction by segmenting them into individual events and repositioning each segment in the correct position according to the TSM factor \cite{nagel2009novel}. 

The noise component has been previously time stretched by randomizing the phase of each signal frame containing noise \cite{roebel2010shape, damskagg2017audio}. However, this leads to an audible disturbance at large time-stretching factors \cite{damskagg2017audio}. This letter proposes to use a morphing technique to time-stretch the noise component with an improved perceptual quality, as described next.

\section{Noise Morphing}\label{s:nm}


This section introduces NM, a spectral morphing technique designed for the independent stretching of the noise component. A similar concept has been explored in previous works of Moinet \cite{moinet2013slowdio} and Apel \cite{apel2014sinusoidality}, although there were small but significant differences. 
The core principle of the NM method revolves around applying random phases while maintaining a magnitude consistent with the original audio, in such a way that perfect correlation between successive STFT frames is ensured.
The proposed approach is grounded in the assumption that the noise or residual component, being quasi-stochastic, has little perceptual impact from its phase, allowing us to discard it.





 

\begin{figure}[t!]
\centering
\includegraphics[width=1\columnwidth,clip]{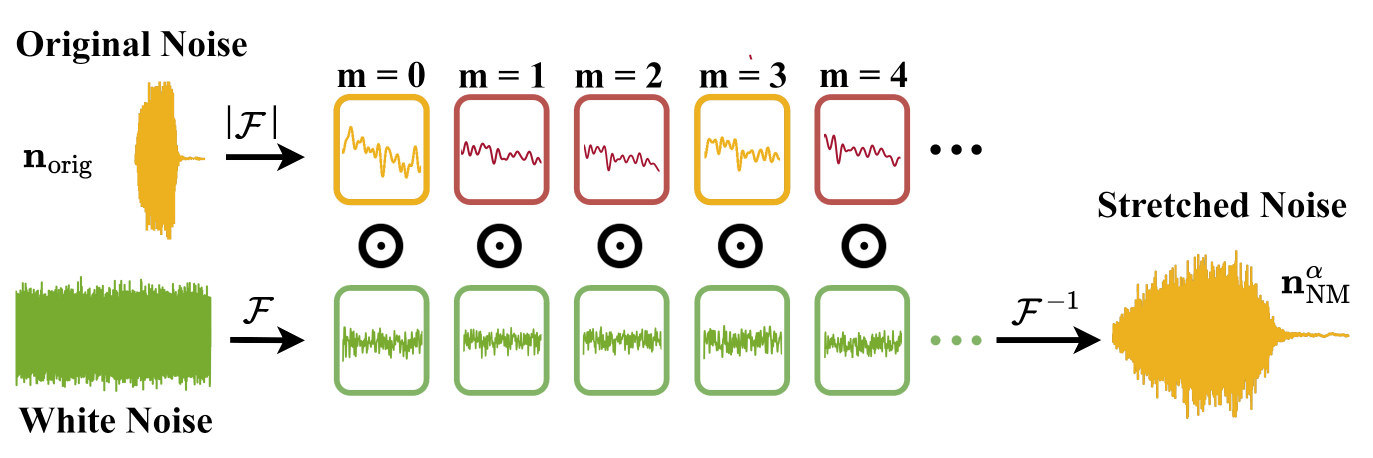}
\caption{Conceptualization of noise morphing, for $\alpha = 3$. The original noise log-magnitude spectra (yellow) are time-interpolated (red) and used to modulate the white-noise spectra (green) to produce the time-stretched output. 
}
\label{fig:noisemorphing}
\end{figure}

The proposed algorithm, depicted in Fig. \ref{fig:noisemorphing}, follows a structured analysis and synthesis procedure. The original noise component $\mathbf{n}_\text{orig} \in \mathbb{R}^N$ is first processed with the STFT, using a Hann window of 2048 samples (46\,ms) and a hop size of 1024 samples (23\,ms) at a sample rate $f_\text{s}=44.1$\,kHz. The log-magnitude spectrum of each STFT frame $\mathbf{N}_\text{orig} \in \mathbb{R}^{M\times K}$ is computed as
\begin{equation}
\mathbf{N}_\text{orig}=10\log_{10}(|\mathcal{F}(\mathbf{n}_\text{orig})|),
\end{equation}
where $\mathcal{F}()$ represents the STFT operator. The log-magnitude spectrum is then linearly interpolated according to the stretching factor $\alpha$ based on the two neighboring spectra, occurring before and after the interpolation point, following
\begin{equation}
    \mathbf{N}^\alpha=\text{lerp}(\mathbf{N}_\text{orig}, \alpha),
\end{equation}
\noindent where $\text{lerp}(\cdot)$ is the linear interpolation function and $\alpha$ is the stretching factor. In the time dimension, the length of the spectrogram $\mathbf{N}^\alpha \in \mathbb{R}^{\alpha M \times K}$ is $\alpha$ times that of the spectrogram $\mathbf{N}_\text{orig} \in \mathbb{R}^{M \times K}$. 

\begin{figure*}[t!]
    \centering
    \includegraphics{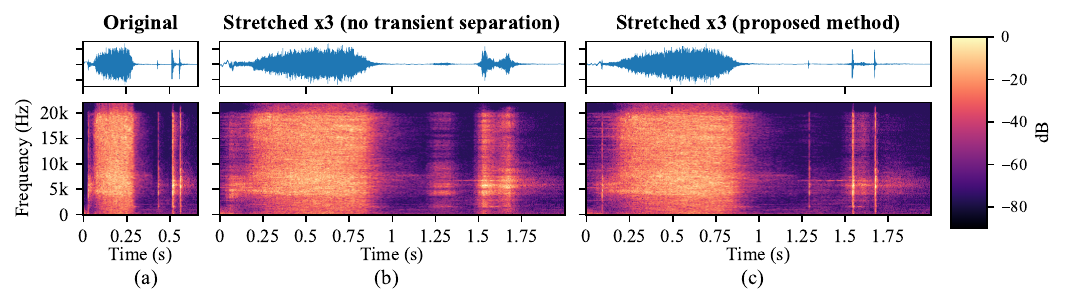}
    \caption{A can opening sound (a) at normal speed and stretched with $\alpha = 3$ (b) without transient separation, which leads to transient smearing, and (c) with the proposed method, which preserves transients with apt handling of the noise component.}   \label{fig:spectrogramspectrogramss}
\end{figure*}

In the synthesis phase, a white-noise excitation signal $\bm{\epsilon} \in \mathbb{R}^{\alpha N}$
is first generated matching the length of the output signal after time stretching, as shown in Fig. \ref{fig:noisemorphing}. According to our experiments, the perceptual impact of the noise sequence's distribution is negligible, provided its spectrum is white, and the sequence is standardized with zero mean and unit variance. Consequently, uniformly or Gaussian distributed noises, when normalized, are both viable options. In this work, the noise signal is sampled from a standard Gaussian distribution.

As shown in Fig.~\ref{fig:noisemorphing}, the STFT is also applied to the white noise, using the same window and hop size as above. 
The resulting complex time-frequency signal $\bm{\mathcal{E}} \in \mathbb{C}^{\alpha T \times F}$ must be normalized by the window energy to ensure that the flat spectral magnitude equals one.
Subsequently, the noise spectral frames are modulated by the interpolated magnitude spectra via element-wise multiplication:
\begin{equation}
    \mathbf{N}_{\text{NM}}^\alpha = \bm{\mathcal{E}} \odot 10^{ \mathbf{N}^\alpha/10}.
    \label{eq:modulation}
\end{equation}

\noindent Finally, the morphed noise signal in the time-domain $\mathbf{n}^\alpha_\text{NM} \in \mathbb{R}^{\alpha N}$ is obtained by applying the inverse STFT using the same parameters as in the analysis (see also Fig.~\ref{fig:noisemorphing}): 
\begin{equation}
    \mathbf{n}^\alpha_\text{NM}=\mathcal{F}^{-1}( \mathbf{N}_\text{NM}^\alpha).
\end{equation}


A notable difference between our method above and the work of Moinet et al. \cite{moinet2013audio} is that the latter directly replaces the magnitude of the time-frequency signal $\bm{\mathcal{E}}$ with the interpolated magnitudes through polar coordinates, neglecting the white-noise magnitude spectra. Our observations suggest that the modulation approach of \eqref{eq:modulation} yields a more organic effect, as the stochastic variations in the magnitude of the white-noise signal contribute to a perceptually smoother and less artifact-prone sound. Apel \cite{apel2014sinusoidality} combines the white-noise spectra and the interpolated magnitude spectra in the same way as here, but in his work, the residual component contains a mixture of noise and transients, which leads to the need for additional spectral smoothing techniques to enhance the sound quality.

A crucial parameter shaping the quality of the synthesized time-stretched audio is the window length. A long window introduces a smoother signal, akin to noise, but comes at the expense of diminished temporal detail in the output signal, and rapidly changing nuances tend to get smeared.
On the contrary, a short window captures finer nuances of the sound, enhancing overall clarity, but has the potential of introducing musical noise artifacts, which may compromise the quality of the synthesized sound. Moinet made
similar observations regarding the window length \cite{moinet2013slowdio}. However, the challenges associated with long windows become more pronounced when transients are not separated. 
Moreover, our approach of multiplying the noise spectral frames with the interpolated magnitude spectra achieves more natural results with a short window, compared to replacing the magnitudes as Moinet suggested \cite{moinet2013audio}. 

\subsection{Audio Time-Stretching Example}

A comprehensive insight into the efficacy of the proposed TSM method is offered by the example visualized in Fig.\;\ref{fig:spectrogramspectrogramss}. The waveform and spectrogram of the unprocessed signal, featuring hisses and clicks from the opening of a soda can, are shown in Fig.~\ref{fig:spectrogramspectrogramss}(a). The stretched noise is highlighted in  Fig.~\ref{fig:spectrogramspectrogramss}(b), as well as the need for transient preservation: when the signal is stretched by a factor of 3, transients between 1.5 and 1.75\,s are clearly smeared over time, resulting in a characteristic undesirable effect. In striking contrast, Fig.~\ref{fig:spectrogramspectrogramss}(c) showcases the proposed method's performance by preserving the transients between 1.5 and 1.75\,s during the time-stretching process. 
Notably, the method adeptly manages the stretching of the noise component appearing around 5\,kHz starting at about 1.5\,s. when transients are separated, emphasizing its ability to achieve desirable audio TSM outcomes.

\section{Evaluation}\label{s:eval}

The proposed method has been evaluated against a set of relevant baselines by means of a formal blind listening test. The evaluation process and results are reported in this section.

\subsection{Compared Methods}  \label{sec:compared_methods}
We considered several baseline methods to provide a comprehensive benchmark for our proposed approach (NM). To establish a lower performance threshold, we included a standard phase vocoder \cite{moulines1995non, timefrequency2011dafx} as anchor (AN). As additional baselines, we incorporated the fuzzy phase vocoder \cite{damskagg2017audio} (FZ) and its enhanced version with transient preservation \cite{fierro2023enhanced} (FT). Furthermore, we integrated a prior method in which the stretching of the noise component was achieved using a neural synthesizer \cite{fierro2023extreme} (WN).

In addition to these baselines, we conducted two ablation studies aimed at elucidating crucial factors influencing the time-stretching quality of the proposed method.
One variant of our approach involved applying noise morphing without prior decomposition and transient separation (ND), resembling previous works by Moinet \cite{moinet2013slowdio} and Apel \cite{apel2014sinusoidality}. 
Lastly, we included a version of our proposed method in which the noise morphing employs spectral magnitude replacement instead of multiplication (NI), as suggested by Moinet \cite{moinet2013slowdio}.


 



\begin{figure*}[t!]
\centering
\begin{subfigure}[t]{0.32\textwidth}
\includegraphics[trim={0 0 180pt 0}, clip, height=110pt]{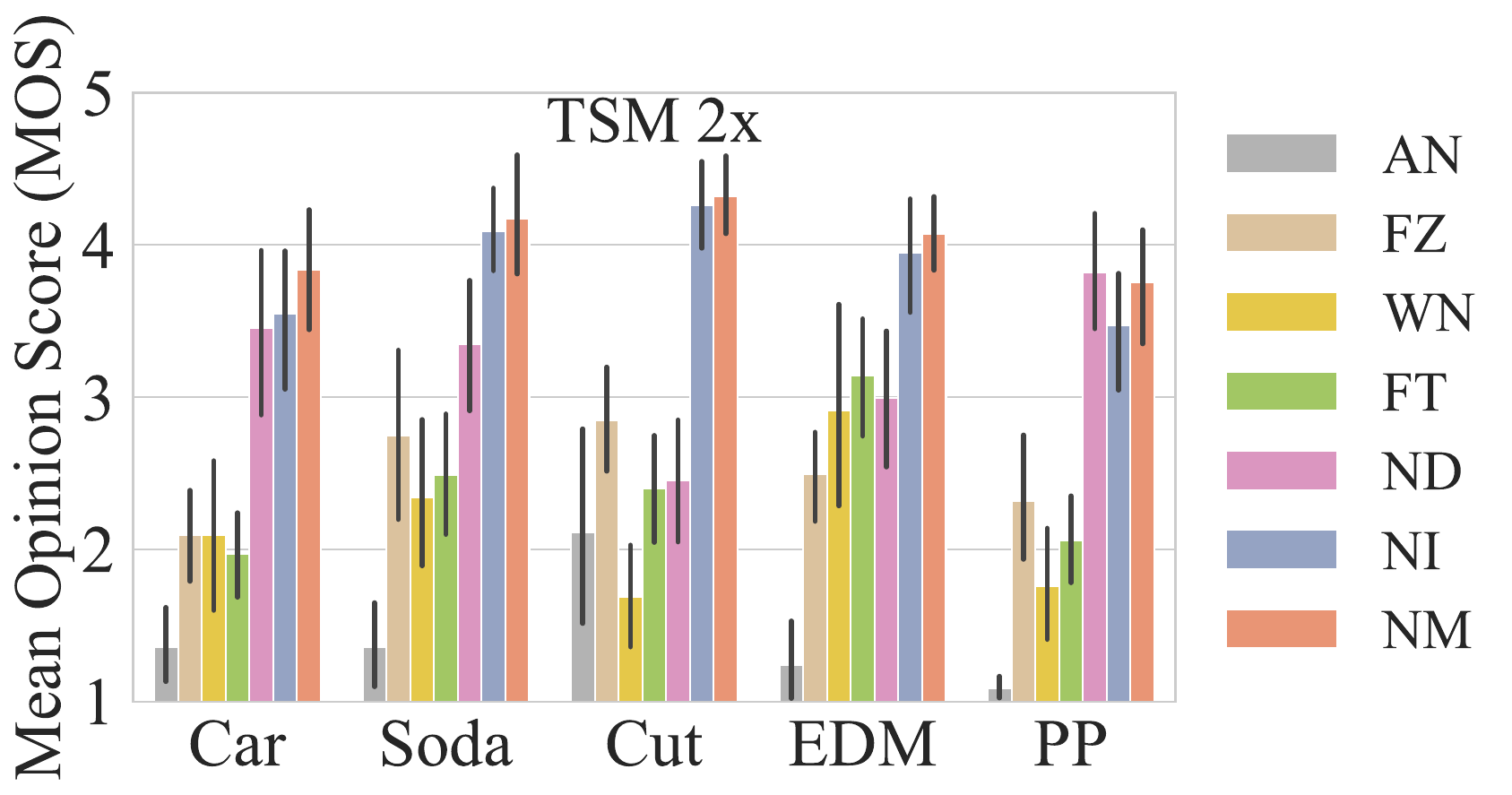}
\caption{}
\end{subfigure}
\hspace{-7pt}
\begin{subfigure}[t]{0.3\textwidth}
\includegraphics[trim={70pt 0 180pt 0}, clip, height=110pt]{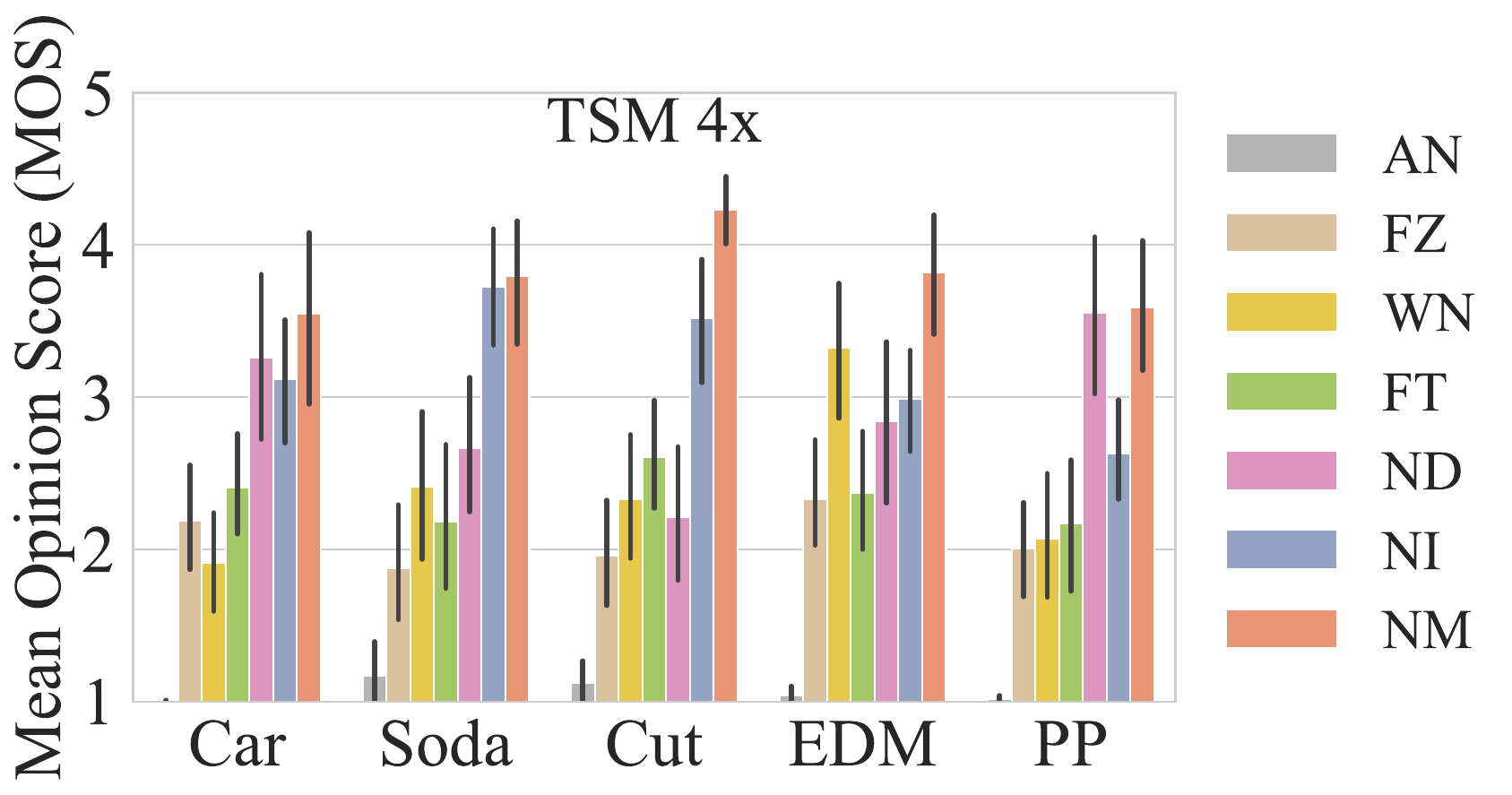}
\caption{}
\end{subfigure}
\hspace{-14pt}
\begin{subfigure}[t]{0.37\textwidth}
\includegraphics[trim={70pt 0 0 0}, clip,  height=110pt]{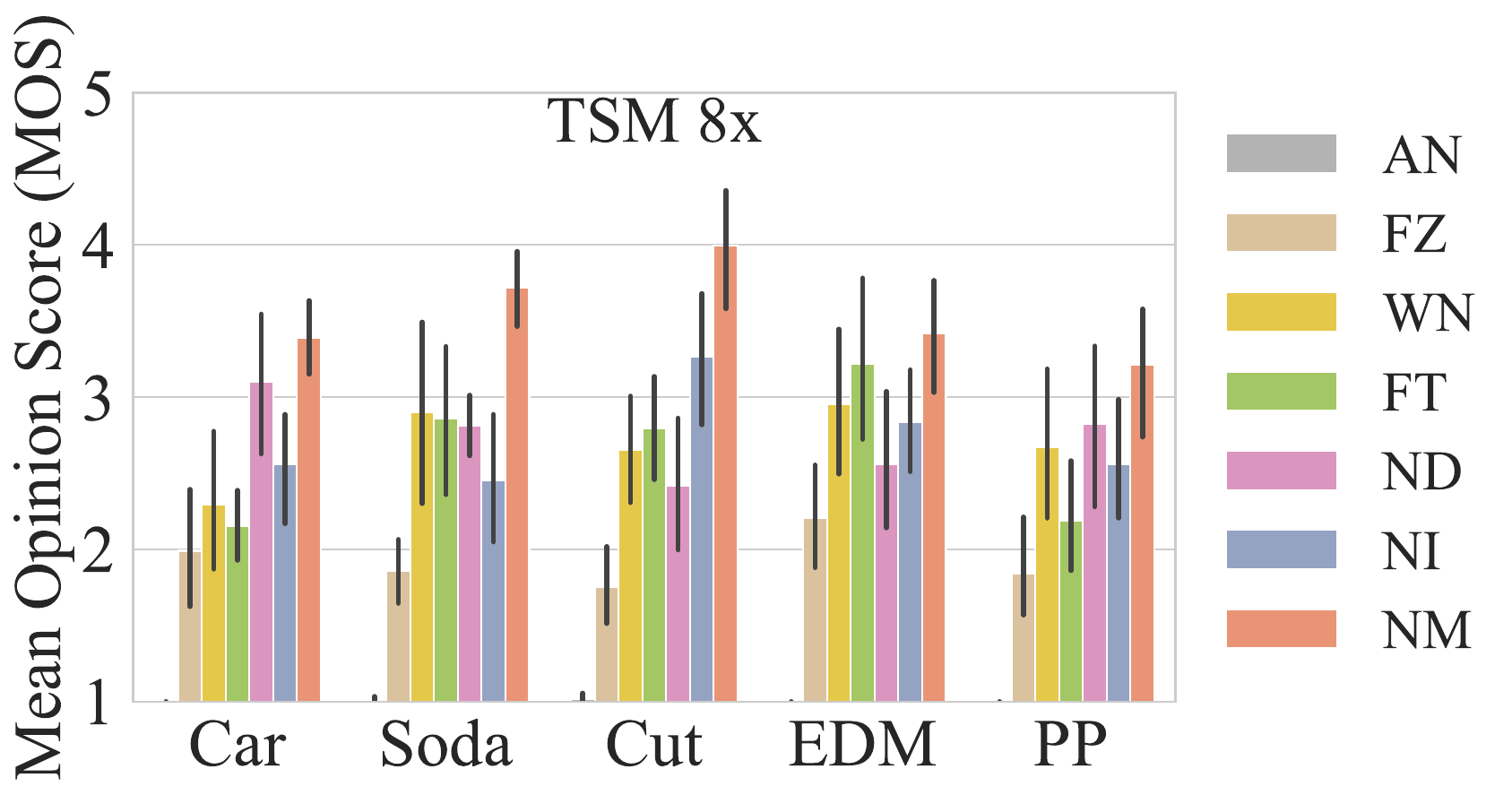}
\caption{}
\end{subfigure}
\caption{Listening test results, showing MOS with 95\% confidence intervals for (a) $\alpha = 2$, (b) $\alpha=4$, and (c) $\alpha=8$.}
\label{fig:eval}
\end{figure*}

\subsection{Listening Test Design}


Our test approach, a variation of the standard MUSHRA test \cite{ITUMUSHRA}, has been used earlier in TSM studies \cite{damskagg2017audio, fierro2023extreme}, and employs a multiple-stimuli method with the original, unprocessed sound as the reference. Across 15 trials, we presented sets of 7 stimuli, with 5 trials being conducted for each TSM factor $\alpha$ = 2,\,4, and \,8. Each set included stimuli representing the proposed method and the 6 baseline methods outlined in Section \ref{sec:compared_methods}. 
A set of 5 representative mono audio excerpts were included in the experiment. 
While we would have preferred to include more examples, we deemed it impractical as it would have resulted in a lengthy and tiring listening test for participants. 
The audio samples under test are listed in Table \ref{tab:examples} and are available on the companion webpage for this letter\footnote{\href{http://research.spa.aalto.fi/publications/papers/ieee-spl-noisemorphing/}{http://research.spa.aalto.fi/publications/papers/ieee-spl-noisemorphing} \label{fn_webpage}}.

\begin{table}[tb!]
\centering
\caption{Audio excerpts used in the listening test}
\begin{tabular}{@{}ll@{}}
\toprule
\textbf{Item name}          & \textbf{Description}         \\ \midrule
Car           & Live recording of a rally car passing by
\\
Soda          & Hiss and click sounds from a can opening \\
Cut           & A knife cutting food on a cutting board   \\
EDM (music)   & Electronic music sample    \\
PP (Ping Pong) & Sounds from an amateur ping pong game   \\ \bottomrule
\end{tabular}%
\label{tab:examples}
\end{table}

To accommodate the extreme stretching factors involved in the test, each audio sample's duration was kept very short (approximately 2\,s). This ensured that the longest time-stretched sounds remained below 18\,s in duration \cite{ITUMUSHRA}.

A total of 13 volunteers participated in the experiment, ranging from 26 to 35 years of age. The participants were instructed to rate each presented stimulus on a scale from 0 to 100, indicating the degree to which the sample met their own subjective expectations for a time-stretched version of the reference, together with the overall audio quality. The participants were not obligated to use the full scale, since ideal examples of best nor worst quality do not exist.

The test software was a customized version of WebMushra \cite{schoeffler2018webmushra}. The audio items were played through a single pair of Sennheiser HD 650 headphones within a soundproof listening booth at the Aalto Acoustics Lab in Espoo, Finland.



\subsection{Results}

The results of the listening test are presented in Fig.~\ref{fig:eval}.
Notably, the proposed Noise Morphing method consistently emerged with the highest Mean Opinion Scores (MOS) across all examples and TSM factors except one, underscoring its efficacy in delivering perceptually superior time-stretched audio. 
The recommended Wilcoxon signed-rank test \cite{mendoncca2018statistical} shows a general trend of statistical significance in the data distributions, despite occasional overlap in some distributions.
Results are reported in the companion website\footref{fn_webpage}.
In this section, our analysis centers on comparing situations where confidence intervals occasionally overlap.


A comparative analysis between NM and NI reveals interesting dynamics. 
For $\alpha$ = 2, NM and NI exhibited similar performance. However, as the stretching factor increased to $\alpha$ = 4 and $\alpha$ = 8, NI received significantly lower scores in most examples. This reinforces our suggestion that the modulation of the magnitude spectra produces a more realistic noise output than simple magnitude replacement. 
Our results indicate that noise morphing without transient decomposition (ND) performs poorly on examples containing clear and frequent transients, such as Cut and EDM.
This observation highlights the beneficial contribution of the STN decomposition in the time-stretching framework.
Interestingly, WN ($\alpha$ = 4) and FT ($\alpha$ = 8) show comparable performance in the EDM example, while NI and ND experience a quality drop. This is most likely due to the nature of the sound, suggesting that WN and FT are more suited for time-stretching music signals.


Qualitative comparisons with Élastique, a renowned piece of commercial software for audio TSM, are not directly addressed here; instead, readers are directed to audio examples available on the accompanying website\footref{fn_webpage} due to the need for third-party software. This limitation precluded a direct quantitative comparison within our controlled testing environment.

To provide an overview of NM capabilities wider than what is shown in the listening test, a larger subset of processed examples is also available for listening on the companion website\footref{fn_webpage}.

\section{Conclusions}\label{s:end}

This letter introduces a method to improve the time-stretching of the noise component of an audio signal, which is obtained by separating tonal and transient components. The proposed Noise Morphing method exhibits consistent superiority in audio quality across various stretch factors when compared to baseline methods. 
The suggested approach shows potential for extensive use in various slow-motion media productions, including music processing or sports videos.
Future work involves exploring 
how to expand the method for stereo and multichannel audio signals.


\clearpage
\bibliographystyle{ieeetr}
\bibliography{refs}

\begin{thebibliography}{10}

\bibitem{moulines1990pitch}
E.~Moulines and F.~Charpentier, ``Pitch-synchronous waveform processing techniques for text-to-speech synthesis using diphones,'' {\em Speech Commun.}, vol.~9, pp.~453--467, Dec. 1990.

\bibitem{bonada2000automatic}
J.~Bonada, ``Automatic technique in frequency domain for near-lossless time-scale modification of audio,'' in {\em Proc. Int. Computer Music Conf.}, (Berlin, Germany), p.~396–399, Aug. 2000.

\bibitem{driedger2016review}
J.~Driedger and M.~M{\"u}ller, ``A review of time-scale modification of music signals,'' {\em Appl. Sci.}, vol.~6, no.~2, p.~57, 2016.

\bibitem{damskagg2017audio}
E.-P. Damsk{\"a}gg and V.~V{\"a}lim{\"a}ki, ``Audio time stretching using fuzzy classification of spectral bins,'' {\em Appl. Sci.}, vol.~7, p.~1293, Dec. 2017.

\bibitem{cliff2000hang}
D.~Cliff, ``Hang the {DJ}: Automatic sequencing and seamless mixing of dance-music tracks,'' {\em HP Lab. Tech. Rep.}, vol.~104, 2000.

\bibitem{Valimaki2018}
V.~V{\"a}lim{\"a}ki, J.~R{\"a}m{\"o}, and F.~Esqueda, ``Creating endless sounds,'' in {\em Proc. 21st Int. Conf. Digital Audio Effects (DAFx)}, (Aveiro, Portugal), pp.~32--39, Sep. 2018.

\bibitem{malloy2022timbral}
C.~Malloy, ``Timbral effects: {T}he {Paulstretch} audio time-stretching algorithm,'' {\em J. Acous. Soc. Am.}, vol.~151, pp.~A158--A158, Apr. 2022.

\bibitem{moinet2013slowdio}
A.~Moinet, {\em Slowdio: Audio Time-Scaling for Slow Motion Sports Videos}.
\newblock PhD thesis, University of Mons, Mons, Belgium, 2013.

\bibitem{roberts2021deep}
T.~Roberts, A.~Nicolson, and K.~K. Paliwal, ``Deep learning-based single-ended quality prediction for time-scale modified audio,'' {\em J. Audio Eng. Soc.}, vol.~69, pp.~644--655, Sept. 2021.

\bibitem{laroche1997phase}
J.~Laroche and M.~Dolson, ``Phase-vocoder: About this phasiness business,'' in {\em Proc. IEEE Workshop Appl. Signal Process. Audio Acoust. (WASPAA)}, (New Paltz, NY), Oct. 1997.

\bibitem{fierro2020towards}
L.~Fierro and V.~V{\"a}lim{\"a}ki, ``Towards objective evaluation of audio time-scale modification methods,'' in {\em Proc. Sound Music Comp. Conf. (SMC)}, (Torino, Italy), pp.~457--462, Jun. 2020.

\bibitem{roebel2010shape}
A.~Röbel, ``A shape-invariant phase vocoder for speech transformation,'' in {\em Proc. 13th Int. Conf. Digital Audio Effects (DAFx-10)}, (Graz, Austria), p.~298–305, Sep. 2010.

\bibitem{driedger2013improving}
J.~Driedger, M.~M{\"u}ller, and S.~Ewert, ``Improving time-scale modification of music signals using harmonic-percussive separation,'' {\em IEEE Signal Process. Lett.}, vol.~21, pp.~105--109, Jan. 2014.

\bibitem{driedger2014tsm}
J.~Driedger and M.~M{\"u}ller, ``{TSM Toolbox: MATLAB} implementations of time-scale modification algorithms,'' in {\em Proc. Int. Conf. Digital Audio Effects (DAFx)}, (Erlangen, Germany), pp.~249--256, Sep. 2014.

\bibitem{roma2019time}
G.~Roma, O.~Green, and P.~A. Tremblay, ``Time scale modification of audio using non-negative matrix factorization,'' in {\em Proc. Int. Conf. Digital Audio Effects (DAFx)}, (Birmingham, UK), Sep. 2019.

\bibitem{liao2012stretching}
W.-H. Liao, A.~Roebel, and A.~W.~Y. Su, ``On stretching {Gaussian} noises with the phase vocoder,'' in {\em Proc. 15th Int. Conf. Digital Audio Effects (DAFx)}, (York, UK), pp.~131--134, Sep. 2012.

\bibitem{fierro2023extreme}
L.~Fierro, A.~Wright, V.~Välimäki, and M.~Hämäläinen, ``Extreme audio time stretching using neural synthesis,'' in {\em Proc. IEEE Int. Conf. Acoust. Speech Signal Process. (ICASSP)}, (Rhodes Island, Greece), pp.~1--5, Jun. 2023.

\bibitem{timefrequency2011dafx}
D.~Arfib, F.~Keiler, U.~Zölzer, V.~Verfaille, and J.~Bonada, ``Time-frequency processing,'' in {\em DAFX: Digital Audio Effects} (U.~Zölzer, ed.), pp.~219--278, Chichester, UK: Wiley, 2nd~ed., 2011.

\bibitem{verma1998analysis}
T.~S. Verma and T.~H. Meng, ``An analysis/synthesis tool for transient signals that allows a flexible sines+transients+noise model for audio,'' in {\em Proc. IEEE Int. Conf. Acoust. Speech Signal Process. (ICASSP)}, vol.~6, (Seattle, WA), pp.~3573--3576, May 1998.

\bibitem{levine1998sines+}
S.~N. Levine and J.~O. Smith~III, ``A sines+transients+noise audio representation for data compression and time/pitch scale modifications,'' in {\em Proc. Audio Eng. Soc. 105th Conv.}, (San Francisco, CA), Sep. 1998.

\bibitem{verma1998time}
T.~S. Verma and T.~H.~Y. Meng, ``Time scale modification using a sines+transients+noise signal model,'' in {\em Proc. Digital Audio Effects Workshop (DAFX’98)}, (Barcelona, Spain), pp.~49--52, Nov. 1998.

\bibitem{fierro2023enhanced}
L.~Fierro and V.~V{\"a}lim{\"a}ki, ``Enhanced fuzzy decomposition of sound into sines, transients, and noise,'' {\em J. Audio Eng. Soc.}, vol.~71, pp.~468--480, Jul. 2023.

\bibitem{serra1990spectral}
X.~Serra and J.~Smith, ``Spectral modeling synthesis: A sound analysis/synthesis system based on a deterministic plus stochastic decomposition,'' {\em Computer Music J.}, vol.~14, no.~4, pp.~12--24, 1990.

\bibitem{hanna2003time}
P.~Hanna and M.~Desainte-Catherine, ``Time scale modification of noises using a spectral and statistical model,'' in {\em Proc. IEEE Int. Conf. Acoust. Speech Signal Process. (ICASSP)}, vol.~6, (Hong Kong, China), pp.~181--184, Apr. 2003.

\bibitem{moinet2013audio}
A.~Moinet, T.~Dutoit, and P.~Latour, ``Audio time-scaling for slow motion sports videos,'' in {\em Proc. Int. Conf. Digital Audio Effects (DAFx)}, (Maynooth, Ireland), pp.~2--5, Sep. 2013.

\bibitem{apel2014sinusoidality}
T.~Apel, ``Sinusoidality analysis and noise synthesis in phase vocoder based timestretching,'' in {\em Proc. Australasian Computer Music Conf.}, (Melbourne, Australia), pp.~7--12, Jul. 2014.

\bibitem{cohen2022speech}
E.~Cohen, F.~Kreuk, and J.~Keshet, ``Speech time-scale modification with {GANs},'' {\em IEEE Signal Process. Lett.}, vol.~29, pp.~1067--1071, Apr. 2022.

\bibitem{verma2000extending}
T.~S. Verma and T.~H.~Y. Meng, ``Extending spectral modeling synthesis with transient modeling synthesis,'' {\em Computer Music J.}, vol.~24, no.~2, pp.~47--59, 2000.

\bibitem{fitzgerald2010harmonic}
D.~Fitzgerald, ``Harmonic/percussive separation using median filtering,'' in {\em Proc. Int. Conf. Digital Audio Effects (DAFx)}, (Graz, Austria), p.~217–220, Sep. 2010.

\bibitem{tachibana2013singing}
H.~Tachibana, N.~Ono, and S.~Sagayama, ``Singing voice enhancement in monaural music signals based on two-stage harmonic/percussive sound separation on multiple resolution spectrograms,'' {\em IEEE Trans. Audio Speech Lang. Process.}, vol.~22, pp.~228--237, Jan. 2014.

\bibitem{driedger2014extending}
J.~Driedger, M.~M{\"u}ller, and S.~Disch, ``Extending harmonic-percussive separation of audio signals,'' in {\em Proc. Int. Conf. Music Inf. Retrieval (ISMIR)}, (Taipei, Taiwan), pp.~611--616, Oct. 2014.

\bibitem{laroche1999improved}
J.~Laroche and M.~Dolson, ``Improved phase vocoder time-scale modification of audio,'' {\em IEEE Trans. Speech Audio Process.}, vol.~7, pp.~323--332, May 1999.

\bibitem{nagel2009novel}
F.~Nagel and A.~Walther, ``A novel transient handling scheme for time stretching algorithms,'' in {\em Proc. Audio Eng. Soc. 127th Conv.}, (New York, NY), Oct. 2009.

\bibitem{moulines1995non}
E.~Moulines and J.~Laroche, ``Non-parametric techniques for pitch-scale and time-scale modification of speech,'' {\em Speech Commun.}, vol.~16, pp.~175--205, Feb. 1995.

\bibitem{ITUMUSHRA}
{IET}, ``{BS}.1534: Method for the subjective assessment of intermediate quality levels of coding systems,'' Recommendation ITU-R BS.1534-1, International Telecommunication Union, Geneva, Switzerland, 2015.

\bibitem{schoeffler2018webmushra}
M.~Schoeffler, S.~Bartoschek, F.-R. St{\"o}ter, {\em et~al.}, ``{WebMUSHRA}---{A} comprehensive framework for web-based listening tests,'' {\em J. Open Research Software}, vol.~6, Feb. 2018.

\bibitem{mendoncca2018statistical}
C.~Mendon{\c{c}}a and S.~Delikaris-Manias, ``Statistical tests with {MUSHRA} data,'' in {\em Proc. 144th Audio Eng. Soc. Conv.}, (Milan, Italy), May 2018.

\end{thebibliography}

\end{document}